\documentstyle[epsf,aps,prb,multicol]{revtex}
\begin{document}
\newcommand{\Pvec}{{\rm\bf P}}
\newcommand{\Evec}{{\rm\bf E}}
\newcommand{\eps}{\epsilon}  
\newcommand{\veps}{\varepsilon}  
\newcommand{\De}{$\Delta$}
\newcommand{\de}{$\delta$}
\newcommand{\mc}{\multicolumn}
\newcommand{\be}{\begin{eqnarray}}
\newcommand{\ee}{\end{eqnarray}}
\newcommand{\einf}{\varepsilon^\infty}
\newcommand{\ez}{\varepsilon^0}  

\draft \title{Effects of macroscopic-polarization built-in 
electrostatic fields in  III-V nitrides multi-quantum-wells}
\author{Vincenzo Fiorentini}
\address{Istituto Nazionale per la Fisica della Materia 
-- Dipartimento di 
 Fisica, Universit\`a di Cagliari, Cagliari, Italy, and \\
 Walter Schottky Institut, Technische Universit\"a{}t
M\"u{}nchen, Garching, Germany}
\author{Fabio Bernardini}
\address{Istituto Nazionale per la Fisica della Materia 
-- Dipartimento di 
 Fisica, Universit\`a di Cagliari, Cagliari, Italy}
\author{Fabio Della Sala, Aldo Di Carlo, and Paolo Lugli}
\address{Istituto Nazionale per la Fisica della Materia 
-- Dipartimento  di Ingegneria Elettronica, 
Universit\`a di Roma ``Tor Vergata'', Roma, Italy}
\date{\today} 
\maketitle

\begin{abstract}
Huge built-in electric fields have been  predicted to exist in 
wurtzite  III-V nitrides  thin films and multilayers.
Such fields originate from heterointerface discontinuities 
of the macroscopic bulk polarization of the nitrides.
Here we discuss the background theory, the role of spontaneous 
polarization in  this context, and  the   practical implications of
built-in polarization fields in nitride nanostructures. To support our
arguments, we present  detailed self-consistent tight-binding
simulations of typical nitride QW structures in which
polarization effects are dominant.  
\end{abstract}


\begin{multicols}{2}
\section{Introduction}

Spontaneous polarization has long been known to take place
in ferroelectrics. On the other hand, its existence in 
semiconductors with sufficiently low crystal symmetry (wurtzite, at
the very least)  has been  generally regarded as of purely theoretical
interest. Recently,  a series of first principles calculations
\cite{noi.piezo,noi.eps,noi.pol}  has reopened this issue for the 
 technologically relevant III-V nitride semiconductors, whose 
natural crystal structure is, in fact, wurtzite. Firstly,
\cite{noi.piezo} it was shown that the nitrides have a very large
spontaneous polarization, as well as  large piezoelectric
coupling constants.  Secondly, \cite{noi.eps,noi.pol}
it was  directly demonstrated  how  polarization actually
manifests itself as electrostatic fields in   nitride multilayers, due
to the polarization charges resulting from  polarization discontinuities
at heterointerfaces. This charge-polarization relation,
counterchecked in actual ab initio calculations, has been
exploited to calculate dielectric constants.\cite{noi.eps}

While piezoelectricity-related properties are largely standard,
spontaneous polarization is to some extent new in semiconductor
physics, to the  point that, so far, the practical importance of
spontaneous polarization in III-V nitrides nanostructures (multi
quantum wells, or MQW's, are the particular focus of this paper) has
been largely overlooked. It is tantalizingly clear to us, however,
that these concepts may lead  to a direct and unambiguous measurement
of the spontaneous polarization in semiconductors, to the  recognition
of its importance in nitride-based nanostructures, and, hopefully, to
its exploitation in device applications. In the hope of accelerating
these processes, in this paper we show how to account for the effects
of spontaneous polarization  in MQW's, and discuss some prototypical
cases and their possible experimental realization.  
To support our arguments, we present simulations of a typical 
AlGaN/GaN MQW where spontaneous and piezoelectric polarizations 
are about equal.

Among the consequences of macroscopic polarization which we will
demonstrate in this paper let us mention the following:
{\it (a)}  the field caused by
the fixed polarization charge, superimposed on the compositional
confinement potential  of the MQW, red-shifts dramatically the transition
energies  and strongly suppresses interband transitions as the well 
thickness increases;   
{\it (b)} the effects of thermal carrier screening are negligible in
typical MQW's, although not in massive samples;  
{\it (c)}  a quasi-flat-band MQW profile can be approximately recovered
(i.e. polarization fields can be screened) only in the presence  of
very high densities of  free carriers, which are appreciably larger
 than those typical of semiconductor laser structures;
{\it (d)} even in the latter case,  transition
probabilities remain considerably smaller than  the ideal flat band
value, and this reduces quantum efficiency; 
{\it (e)} once an
appropriate screening density (i.e. the pumping power) has been
chosen to ensure that the recombination rate is sufficient, a
residual polarization fields is typically still present: this
 provides a means to intentionally red-shift transition energies by
changing well  thicknesses, without changing composition; 
{\it (f)} the
very  existence of distinct and separately controllable spontaneous
and piezoelectric polarization components allows to choose a
composition  such that they cancel each other out, leading to very
nearly ideal flat-band conditions.  A fuller understanding of these points
cannot be but helpful to the design of real nitride devices, and their
operation, involving carrier generation  by light, current injection,
or doping.

\section{Piezoelectric fields}
Piezoelectricity is a well known concept in semiconductor
physics. Binary compounds of strategic technological importance like
III-V  arsenides and phosphides can be forced to exhibit piezoelectric
polarization fields by imposing upon them a strain field.

Among others, applications of piezoelectric effects in semiconductor 
nanotechnology exist in the area of multi quantum wells (MQW) devices.
A thin semicondutor layer (active layer) is embedded in a
semiconductor matrix (cladding layers) having a different lattice
constant. 
If pseudomophic growth occurs, the active layer will be 
 strained and therefore subjected to a piezoelectric
polarization   field, which can be computed as
\be \Pvec^{\rm (pz)} =  \tensor{e} \cdot \vec{\epsilon}\, \ee
if the strain field $\vec{\epsilon}$ 
and the piezoelectric constants tensor  $\tensor{e}$ 
 are known.

The existence of a polarization field in a finite system
 implies the presence 
of electric fields. For the piezoelectric case, the magnitude of the 
latter  depends on strain, piezoelectric constants , and (crucially) on
  device geometry.
The structure of a typical III-V nitride-based superlattice or
MQW is  --C--A--C--A--C--A--C--
(A=active, C=cladding), where both the cladding layer  and the active
layer  are in general strained to comply with the substrate
in-plane lattice parameter. 
In such a structure,
the  electric fields in the A and C layers are
\be 
\Evec^{\rm (pz)}_{\rm A} =
 4\pi \ell_{\rm C} (\Pvec^{\rm (pz)}_{\rm C}
 - \Pvec^{\rm (pz)}_{\rm A})\, /(\ell_{\rm C}\,\veps_{\rm A} 
 + \ell_{\rm A}\,\veps_{\rm C}) \nonumber
\\
    \Evec^{\rm (pz)}_{\rm C} =
 4\pi \ell_{\rm A} (\Pvec^{\rm (pz)}_{\rm A}
 - \Pvec^{\rm (pz)}_{\rm C})\, /(\ell_{\rm C}\,\veps_{\rm A} 
 + \ell_{\rm A}\,\veps_{\rm C})
\label{eq.piezo}
\ee
where $\veps_{\rm A,C}$ are the dielectric constants  and $\ell_{A,C}$ 
the thicknesses of  layers A and C. Thus, in general, an electric
field will be present whenever  $\Pvec_{\rm A} \neq \Pvec_{\rm
C}$. The above expressions are easily obtained \cite{nota3} by the
conditions that the electric displacement be conserved along the
growth axis, and by the boundary conditions that the potential energy
on the far right and left of the MQW structure are the
same. \cite{nota1} 

There are essentially three special cases of MQW structures  worth a
mention: 
\begin{itemize}
\item[~~i)] 
active (cladding) layer lattice matched to the substrate:  $\Pvec_{\rm
A}= 0$ ($\Pvec_{\rm C}= 0$); \item[~ii)]
$\ell_{\rm A} = \ell_{\rm C}$, whence $\Evec_{\rm A} = -\Evec_{\rm C}$;
\item[iii)] $\ell_{\rm A}  \ll \ell_{\rm C}$, implying  $\Evec_{\rm C}
\simeq 0$, and  hence
\end{itemize}
\be \Evec^{\rm (pz)}_{\rm A} = 4\pi \Pvec^{\rm (pz)}_{\rm A} /\veps_{\rm A} 
.\ee
In the last case  we implicitly assumed the cladding layer
to be unstrained --
that is, its lattice constant to be relaxed to its equilibrium value 
because its thickness exceeds the critical value for pseudomorphic
growth  over the substrate. 
 $\Pvec^{\rm (pz)}$ may take any direction in general, but in 
normal practice its direction is parallel to the growth 
axis. \cite{nota}

To obtain piezoelectric  polarization effects in
zincblende semiconductor systems, lattice-mismatched epitaxial 
layers are purposely grown  along a polar axis, e.g. (111); the
in-plane strain propagates elastically onto the growth direction,
thereby generating $\Pvec^{\rm (pz)}$. In wurtzite nitrides, the
preferred growth direction is the polar (0001) [or (000$\overline{\rm
1}$)] axis, so that any non-accomodated in-plane mismatch automatically
generates a piezoelectric polarization along
 the growth axis (the sign depends on whether the epitaxial strain is 
compressive or tensile). We will be always be assuming this situation in
 the following. 

Usually, alloys are employed in the fabrication of MQW's. In that case,
one may estimate the piezoelectric polarization in the spirit of
Vegard's  law as, for a general strain imposed upon e.g. an
Al$_x$Ga$_{1-x}$N  alloy,
\be
    \Pvec^{\rm (pz)} 
= \left[ x ~\tensor{e}_{\rm AlN} + (1-x) ~\tensor{e}_{\rm
       GaN} \right]  	    ~\vec{\epsilon} \,(x)\, ,
\ee
This expression  contains terms linear as well as quadratic in $x$,
and similar relations hold for quaternary solutions. This piezoelectric
 term is only present in pseudomorphic strained growth, and will typically
tend to zero beyond the critical thickness  at which strain relaxation
sets in. Uncomfortable though it may be,\cite{zunger} the Vegard hypotesis is at
this point in time the only way we have to account for piezoelectric
(and spontaneous, see below) fields in alloys. As will be shown below,
indeed, the qualitative picture does not depend so much on the detailed
value of the polarization field as on their order of magnitude.

\section{Spontaneous  fields in MQW's}
New possibilities are opened by the use of III-V nitrides
(AlN,GaN,InN), that naturally cristallize in the wurtzite structure.
These materials are characterized by polarization properties that
differ dramatically from those of the standard III-V compounds
considered so far. From simple symmetry arguments,\cite{nye} it can
be shown that  wurtzite  semiconductors are characterized  by 
 a non-zero polarization in their equilibrium (unstrained) geometry,
named spontaneous polarization (or, occasionally, pyroelectric, with
reference to its change with temperature).\cite{nye2}
While the spontaneous polarization of ferroelectrics can  be  measured
via an hysteresis cycle, in a wurtzite this cannot be done, 
since no hysteresis can take place in that structure.
Indeed,  spontaneous polarization has never been measured in wurtzites
so far.  III-V nitrides MQW's offer  the opportunity to reveal
its existence and to actually measure it. In turn, spontaneous
polarization can provide new degrees of freedom, in the form of
permanent {\it strain-independent}
 built-in electrostatic fields, to tailor transport and
optical characteristics of nitride nanostructures.  
 Its presence can e.g.  be exploited to cancel out the
piezoelectric fields produced in typical  pseudomorphically strained
nitride structures, as discussed below.

Thanks to recent advances\cite{KS} in the modern theory of polarization
(a unified approach based on the Berry's phase concept),
it has become possible to compute easily and accurately from first
principles the values of the spontaneous polarization, besides
piezoelectric and dielectric  constants, in III-V
nitrides.\cite{noi.piezo,noi.eps} The results of the calculations
show that III-V nitrides  have important
polarization-related properties that set them apart from 
{standard} zincblende III-V semiconductors: 
\begin{itemize}
\item[~~i)]
huge piezoelectric constants (much larger than, and opposite in sign
to  those of all other III-V's);
\item[~ii)] existence of a spontaneous polarization field of the 
same order of magnitude as in ferroelectrics.
\end{itemize}
The latter is, we think, a most relevant property. Spontaneous
polarization implies  that even in heterostructure systems where
active and cladding layers are both lattice-matched to the substrate 
(so that no strain occurs, and hence
no piezoelectricity), an electric field will nevertheless exist  due
to spontaneous polarization.
In addition, unlike piezoelectric polarization, spontaneous
polarization has a {\it fixed direction} in the crystal: in wurtzites
it is the (0001) axis, which is (as mentioned previously) the growth
direction of choice for nitrides epitaxy.  Therefore the field
resulting from spontaneous polarization will point along the growth  
direction, and this  {\it (a)} maximizes spontaneous
 polarization effects in these systems, and {\it  (b)} it renders the
problem effectively one-dimensional.
In the simplest case of a fully unstrained (substrate lattice-matched)
MQW,  the electric  fields inside the layers are given, in analogy to
Eq. \ref{eq.piezo}, by
\be
\Evec^{\rm (sp)}_{\rm A} = 4\pi \ell_{\rm C} (\Pvec^{\rm (sp)}_{\rm C} 
- \Pvec^{\rm (sp)}_{\rm A})  
		       /(\ell_{\rm C}\veps_{\rm A} + \ell_{\rm
A}\veps_{\rm C}) \nonumber   
   \\
\Evec^{\rm (sp)}_{\rm C} = 4\pi \ell_{\rm A} (\Pvec^{\rm (sp)}_{\rm A} 
- \Pvec^{\rm (sp)}_{\rm C}) 
		       /(\ell_{\rm C}\veps_{\rm A} + \ell_{\rm A}\veps_{\rm C})
\label{eq.spont}
\ee
where the superscript ${\rm (sp)}$ stands for spontaneous;
 typical spontaneous  polarization values\cite{noi.piezo}
indicate that these fields are very large 
(up to several MV/cm).

\begin{figure}[h]
\epsfclipon
\epsfxsize=8cm
\centerline{\epsffile{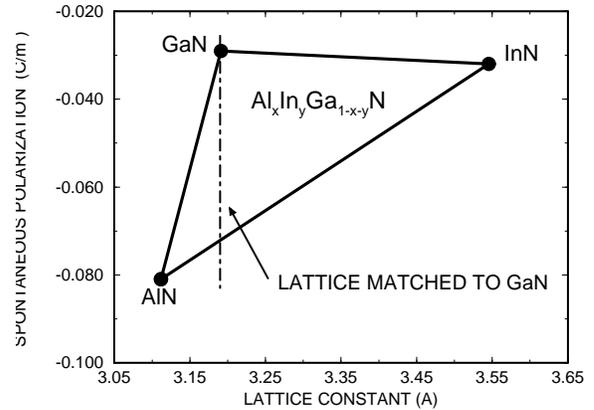}}
\caption{Spontaneous polarization in Al$_x$In$_y$Ga$_{1-x-y}$N
 alloys according to
a Vegard-like rule.}
\label{fig.vegard}
\end{figure}

 In actual applications  (for instance, to produce unstrained
MQW's)  alloys will have to be employed. The
values of the spontaneous polarization are accurately known only for
binary compounds.\cite{noi.piezo} In the absence of better estimates,
we assume as before that the spontaneous polarization in alloys can be
estimated  using a Vegard-like rule as
(for, e.g., Al$_x$In$_y$Ga$_{1-x-y}$N)
$$    \Pvec^{\rm (sp)}(x,y) =
 x ~\Pvec^{\rm (sp)}_{\rm AlN} + y ~\Pvec^{\rm (sp)}_{\rm InN} 
                         + (1-x-y) ~\Pvec^{\rm (sp)}_{\rm GaN}\, .
$$
 In Fig. \ref{fig.vegard} we report
the resulting  spontaneous polarization vs. lattice constant for the  
III-V nitrides,  with data   from Ref.\onlinecite{noi.piezo}.

\narrowtext
Fig. \ref{fig.vegard} shows that for a given (substrate) lattice
constant, a  wide interval of spontaneous polarizations (hence of
spontaneous fields, according to Eq. \ref{eq.spont}) is accessible
varying   the alloy composition. In particular, consider a
GaN/Al$_x$In$_{y}$Ga$_{1-x-y}$N MQW,  where the composition is 
chosen so that the alloy be lattice matched to GaN, which we assume to
be also the substrate (or buffer) material (dashed-dotted line in
Fig. \ref{fig.vegard}). Then, piezoelectric polarization vanishes,
but spontaneous polarization remains, and takes on values up to
$\sim$0.05 C/m$^2$. For a GaN quantum well with very thick AlGaN
cladding layers, this means a theoretical electrostatic field of 
up to about 5 MV/cm.

\section{Fields in the general case}
 In general, of course, MQW's will be strained.
 Then,  for an arbitrary strain state, the electric fields 
in the A (or C) layers of the MQW are {\it the sum} of the 
piezoelectric and spontaneous
contributions: 
$$\Evec_{\rm A,C} = \Evec^{\rm (sp)}_{\rm A,C} + \Evec^{\rm (pz)}_{\rm
A,C},$$ 
where $\Evec^{\rm (pz)}$ is the old-fashioned piezoelectric field from
Eq.~\ref{eq.piezo}, and $\Evec^{\rm (sp)}$ is given by
Eq.~\ref{eq.spont}. It is important to stress  that 
$\Evec^{\rm (sp)}$ depends only on  material composition and not on
the strain state. Also, it is a key point to notice that although
both polarization contributions lay along the same direction
[the (0001) axis],   $\Pvec^{\rm (pz)}$  may have (due to its
strain dependence) the same or the
opposite sign with respect to the fixed  $\Pvec^{\rm (sp)}$ 
depending on the epitaxial relations.

\narrowtext
\begin{figure}[t]
\epsfysize=7cm
\centerline{\epsffile{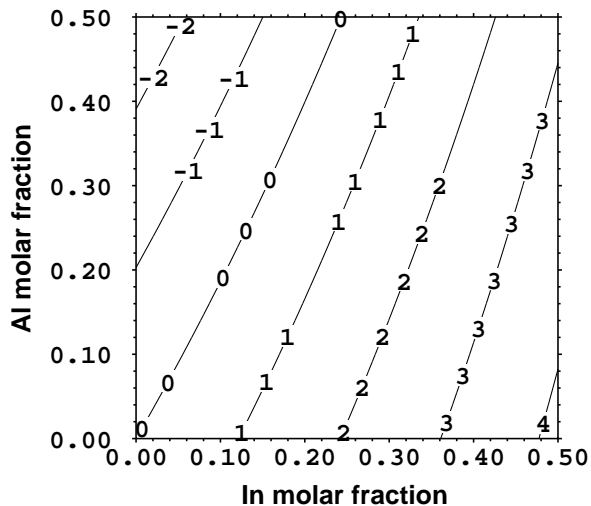}}
\caption{Total built-in electrostatic field in the active layer of a
  Al$_x$In$_y$Ga$_{1-x-y}$N/GaN 
MQW system (see text) vs. Al and In molar fraction.
Fields are in units of MV/cm, positive fields point 
 in the (0001) direction (Ga-face).}
\label{fig.efield}
\end{figure}

\begin{figure}[h]
\epsfysize=7cm\centerline{\epsffile{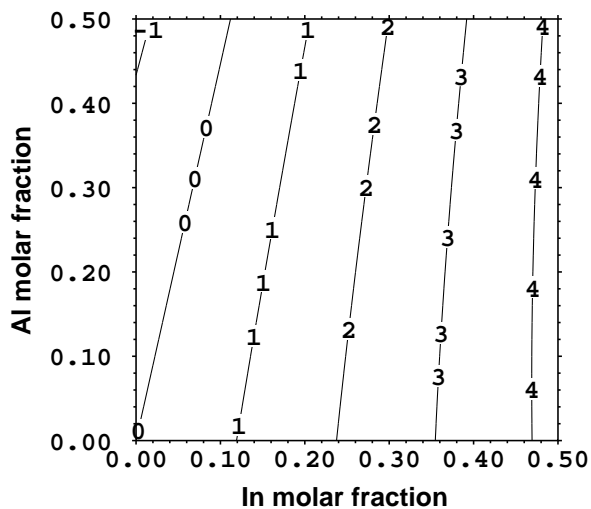}}
\caption{Piezoelectric component of the electrostatic fields
in the same MQW system as  in Fig.\protect\ref{fig.efield} (see text).}
\label{fig.piezo}
\end{figure}

It is difficult to give a simple picture of the
electric field pattern in a general MQW system because of the many
degrees of freedom involved.  Here we consider an
Al$_{x}$Ga$_{y}$In$_{1-x-y}$N/GaN   MQW pseudomorphically grown over a
GaN  substrate, having active and cladding layers of the same thickness.
In such a case 
$$ \Evec^{\rm (sp)}_{\rm A} + \Evec^{\rm (pz)}_{\rm A} \equiv
\Evec_{\rm A} = - \Evec_{\rm C} \equiv -(\Evec^{\rm (sp)}_{\rm C} +
\Evec^{\rm (pz)}_{\rm C})\, . $$  
Note again, at his point, that the fields
(see Eqs. \ref{eq.piezo} and \ref{eq.spont}) are not related to just
the polarization of the material composing the specific layer, but a
combination of polarization {\it differences}, dielectric screening,
 and geometrical factors. \cite{nardelli} We now consider the
field values in  the active layer: the total field $\Evec_{\rm A}$ 
is shown in Fig.~\ref{fig.efield} vs. Al and In molar  fraction; the
same is done for the piezoelectric  component  in
Fig.~\ref{fig.piezo}. In both cases the appropriate Vegard-like rules
have been used. 
\narrowtext
Comparison of these Figures cleary bears out the importance of
spontaneous polarization in determining the electric field.
 Several aspects  are worth pointing out. First, large electric
fields ($\sim$ 0.5--1 MV/cm), can be obtained already   for modest Al
and In concentrations. Second,  it is easy  to access compositions
such that Al$_x$In$_y$Ga$_{1-x-y}$N  is  lattice matched to GaN:
thereby, no piezoelectric fields exist, but  large, purely spontaneous
fields still do; specifically, this situation is realized for
compositions laying on the zero-piezoelectric-field line in
Fig. \ref{fig.piezo}.  Third, it is possible to choose the material
composition  in such a way that the active layers of a 
MQW's are free of electric fields, despite 
the spontaneous polarization. To achieve this situation the MQW must
be strained so that the piezoelectric and 
spontaneous polarizations cancel each other out; clearly, this is
realized for compositions laying on the zero-field line in
Fig. \ref{fig.efield}. 
Of course the possibility of having a null field where desired
 is of capital importance in 
those devices  where electric fields in the active layer can not be
tolerated (other field screening mechanisms  are discussed
below).

 A noticeable feature of 
Fig. \ref{fig.piezo} is that the piezoelectric component increases 
much faster with In content that with Al content, despite the larger
piezoelectric constants of the latter. The reason is, of course,
that strain builds up much more rapidly with In concentration
(in the Vegard hypotesis). 
On the other hand, it can be  seen that the {\it spontaneous}
component increases much more rapidly with Al content than with In
content, due to the widely different polarizations of AlN and GaN.

\section{Effects of polarization fields}
\label{screening}

We now come to the implications of polarization fields for devices
based on III-V nitrides.
 In this Section (with exception of Sec.\ref{sec.mass}) we present 
a set of 
accurate self-consistent tight-binding  calculations for an isolated AlGaN/GaN
QWs representing a system in which the contribution to the total built-in 
electrostatic field of the 
spontaneous polarization 
is as large as the piezoelectric term. 
In realistic simulations of devices, 
self-consistency is needed to describe field screening by free
carriers; the latter cannot physically cancel out the polarization
charge, which is fixed  and invariable, but may 
screen it out in part. In our calculations we therefore  solve
self-consistently the Poisson equation and  
the Schr\"odinger equation for a state-of-the-art
empirical tight binding Hamiltonian   for  realistic nanostructures.
\cite{aldo}
In the following two cases are considered: {\it (a)} non-equibrium
carrier distribution (Sec. A and B) related to photoexcitation or injection,  
where electron and hole quasi-Fermi levels are calculated  for a given
areal charge density ($n_{2D}$) in the quantum well (the {sheet density},
 related to  the injection current or optical pumping power); 
{\it (b)}  thermal equilibrium distribution 
(Sec. D) where the Fermi level is calculated as a function of 
doping density by imposing charge neutrality conditions.\cite{aldo}  
We solve Poisson's equation,
\begin{equation} 
\frac{d}{dz}D =
\frac{d}{dz}\left(-\varepsilon\frac{d}{dz}V+P_T\right)=
e\left(p-n\right),
\label{eq:1}
\end{equation}
where the (position-dependent) quantities  $D$,   $\varepsilon$, and $V$,
 are respectively the displacement field, dielectric constant, and
potential. $P_T$ is the (position-dependent) total transverse
polarization. The effects of composition, polarization, and free
carrier screening are thus included in full. 
Consistently with the aim of describing a {\it single} QW, we choose
the boundary conditions of  zero field at the ends of the simulation
region. 

The potential thus obtained
is inserted in the Schr\"odinger equation,  which is solved
diagonalizing the empirical tight-binding $sp^3d^5s^*$
Hamiltonian. \cite{jancu} The procedure is iterated to self-consistency.
Further applications and details on the
technique can be found  elsewhere.  \cite{aldo,ganapl} 

Here we concentrate in particular on the
polarization-induced quantum-confined Stark effect 
(QCSE) in zero
external field, and its control and
quenching. We first deal with the low free-carrier densities regime:
thereby the QCSE  manifests
itself as a strong red shift of the interband transition energy, with a
concurrent suppression of the transition probability, both of these
features getting stronger as the well thickness  increases. This is
the regime that applies to low-power operation or
 photoluminescence experiments. 

Next we discuss how the QCSE can be modified, and eventually (almost)
quenched, by providing  the QW with a sufficiently high free-carrier
density. In this regime, as the free carrier density increases,  the
transition energy is progressively blue-shifted back towards its flat
band value, and the transition probability suppression is largely
removed. The needed free-carrier density depends on the polarization
field, and not surprisingly it is found to be typically very
substantial. Typical values of the sheet density range in the
10$^{13}$ cm$^{-2}$, as opposed to typical values of
10$^{12}$  cm$^{-2}$ needed to obtain lasing in GaAs-like materials.

\subsection{QCSE at low power}

\begin{figure}[ht]
\epsfclipon
\epsfysize=8cm
\centerline{\epsffile{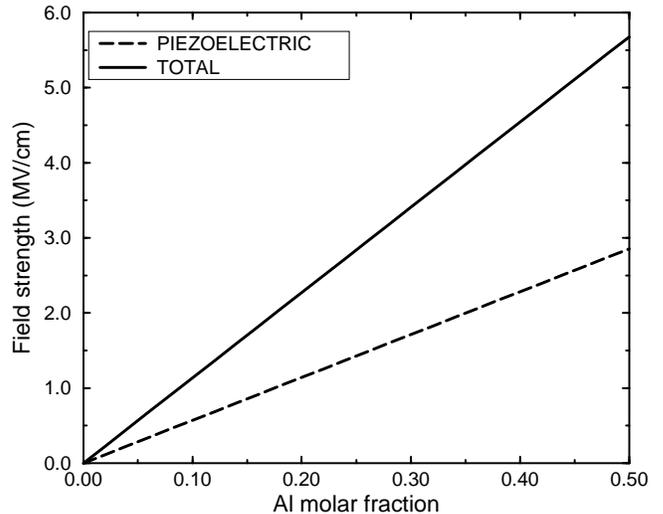}}
\caption{Total field and its piezoelectric component in
the GaN/Al$_x$Ga$_{1-x}$N QW discussed in the text.}
\label{fig.piezo.2}
\end{figure}

The prototypical system we consider is an isolated  GaN 
quantum well cladded between  Al$_{x}$Ga$_{1-x}$N
barriers. In Fig. \ref{fig.piezo.2}, similarly to
Fig. \ref{fig.piezo}, we display the
total field {\bf E}$_{\rm A}$ in the (isolated) active well, and its
piezoelectric component as a function of the
Al molar fraction $x$. The spontaneous component is the difference of
the two, and therefore approximately equal  to
 the piezoelectric one.\cite{nota2}

The value we pick for our simulations is  $x$=0.2, a
reasonable compromise between the conflicting needs for not-too-large fields,
 sufficient  confinement,\cite{noi.pol}
 and technologically achievable composition.
In this case the valence offset is  $\Delta E_v=0.064$ eV.
The total field in the QW of --2.26 MV/cm, and the spontaneous and
piezoelectric components are --1.14 MV/cm and --1.12 MV/cm
respectively. The minus  signs indicates that the field points
in the  (000$\overline{1}$) direction. The bare polarization charge
at the interface is  proportional to the change in polarization
 across the interfaces, and it
amounts to $\sim$1.28$\times$10$^{13}$ cm$^{-2}$. The {\it field}
value mentioned above  results from this charge as screened by the
dielectric response of the QW (the field change at the interface is
thus related to a smaller, or screened, effective interface charge\cite{noi.pol}). 

We have performed a series of calculation for different well width where the
electron and  hole confined states have been populated (i.e. pairs
have been created) with a density of $\sim$
10$^{11}$ cm$^{-2}$ to simulate a low-power optical excitation.
We find  that this density has only a  very marginal effect:
indeed, the potential is perfectly  linear, i.e. the electrostatic field
remains uniform, over essentially all the QW.
The square-to-triangular change in the potential shape causes a small
blue shift 
of both the electron and hole confined states (referred to the flat 
well bottom), but the linear potential
given by the field causes a much larger relative red shift for any reasonable
thickness. Also, since the thermal carrier density fluctuations are 
negligible at microscopic thicknesses and room temperature (see below
and Ref.\onlinecite{mermin}), one expects  the QW band edge profile
to  remain linear as function of thickness, at least for the low
excitation powers  typical of photoluminescence spectroscopy.     
 
\begin{figure}
\epsfclipon
\epsfxsize=7.5cm
\centerline{\epsffile{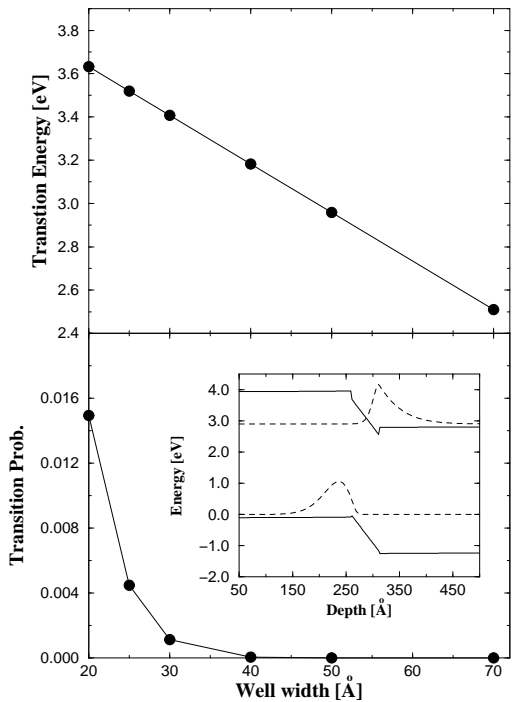}}
\caption{Transition energy red shift and suppression of transition probability
vs well thickness. In the inset self-consistent band edge (solid), 
electron and hole envelope functions of the TB wavefunctions (dashed) for 
a 50 \AA ~thick QW.
}
\label{fig.low.2}
\end{figure}

In Figure \ref{fig.low.2}
we show the TB result for the lowest interband transition energy and the 
corresponding  transition probability 
(i.e. the  squared overlap 
of the highest
hole level and the lowest electron level envelope wavefunctions~\cite{aldo}) 
as a function of QW
thickness. 
Both the Stark red shift and the strong suppression 
of the transition probability are evident, as was to be expected 
from the potential shape and the reduced overlap of hole and electron 
states (see inset Fig. 5).

It is worth noting that the localization of the hole envelope function in
 the well region is rather weak, because 
the large effective field blue-shifts the hole bound state energy 
close to  the valence barrier edge. This will generally be the case for
 low-$x$ AlGaN wells, due to  the small valence confinement 
energy.\cite{noi.pol} In
 fact, on the scale of the  fields-induced potential drop, even the
 conduction confinement is  small, and the electron bound state also tends to
 have the character of a  resonance for small $x$ (i.e. small confinement).

We conclude that  in the absence
of excitation and at normal operation temperatures, or 
at low optical excitation powers, macroscopic
polarization fields  cause QW's to be highly inefficient in 
emitting light, and the emission energy to be 
considerably different than the gap of the material plus confinement
energies. 

Comparison with experiment is tricky since most attempts to measure
these effects are polluted by (typically unnecessary) complicated choices of
the geometry. In any case, the general experimental features
\cite{takeuchi,nak} are in full agreement with the notion that the
transitions are red-shifted essentially linearly with increasing well
thickness, and that screening at low free carrier densities is
irrelevant in this class of systems. This is not quite true any more
for thick layers, as will be discussed in Sec.\ref{sec.mass}.   

\subsection{QCSE quenching at high excitation power}

If  carriers are generated optically, one can envisage  that a
sufficiently high excitation power could possibly produce the carrier
density needed to screen the polarization field. We now calculate the
properties of the QW as a function of the  free-carrier areal density,
to check if the red shift and the  transition probability suppression
can be removed in a physically accessible range of such density.     

\begin{figure}
\epsfclipon\epsfysize=9cm\centerline{\epsffile{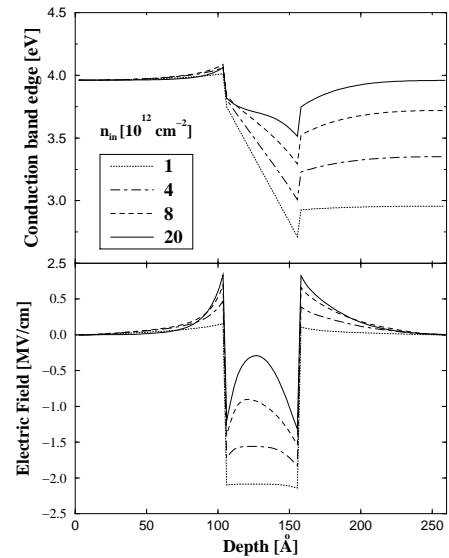}}
\caption{Self-consistent band edge (a) and electric field (b) 
for various sheet densities ($\sim$ excitation level) in a 50 \AA ~thick
QW.}
\label{fig.high.1}
\end{figure}

\begin{figure}
\epsfclipon
\epsfxsize=7cm
\epsfysize=7.5cm\centerline{\epsffile{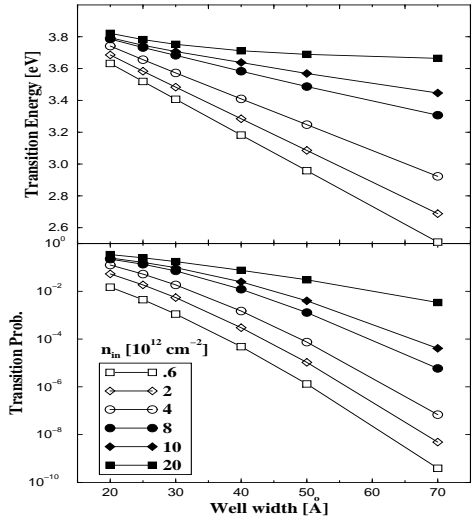}}
\caption{Removal of red shift and recovery of interband transition probability
upon high excitation.}
\label{fig.high.2}
\end{figure}

We repeat the self-consistent procedure  increasing progressively the
free charge density in the QW, from  10$^{12}$ up to 2 $\times$
10$^{13}$ cm$^{-2}$.  We see in Fig. \ref{fig.high.1} that,
albeit  at the cost of a large increase of the QW free-carrier
density, the field does get progressively screened.
As can be
seen in Fig. \ref{fig.high.2}, at fixed thickness the  red shift
decreases as a function of  carrier density,
and it tends to become thickness-independent at the highest densities.
The transition probability is also increased by several orders of magnitude;
however, the field is not screened abruptly but 
dies off gradually, with an effective screening length of about 20
{\AA} for the largest density used here (of course, this is a token of
the larger spatial extension of the screening charge as compared to
the polarization charge \cite{noi.pol}). Therefore, holes and electron
remain spatially separated to a large extent even at high carrier
densities, and the trasition probability never quite  goes back to
unity. [This is not unlikely to be one of the reasons  for the
relatively low quantum efficiency observed in typical nitride MQW
devices.] For the same reasons, the  transition energy never goes back
exactly to the flat-band value (gap  plus confinement energy). Note in
passing that because of strain,   in these calculations  $E_g^{\rm
GaN} = 3.71$ eV, almost 10 \% larger than the equilibrium value.

The screening density  of order 2 $\times 10^{13}$ cm$^{-2}$ needed
to partially screen out the field   corresponds to an optical pumping power of
about 10 to 20 kW/cm$^2$ per well, as can be estimated from 
Refs.\onlinecite{niwa,nomura,radtke}. This figure 
  agrees nicely with the unusually  high
pumping powers\cite{nak,domen} needed to obtain the laser effect in nitride
structures. The explanation is simply that much of the free
 charge being generated actually goes into screening the polarization    
field.  On the other hand, our result prove that the optically activated 
lasing conditions can indeed be realized in practice, although with
high pumping powers, so that there seems to be  no need to invoke
quantum dot  formation\cite{nardelli} or other exotic effects to
explain lasing in nitride structures. On the other hand, the same
phenomenon  explains the high current threshold observed for
electrically driven GaN  based lasers.\cite{nakamura:2,fang:1,yeo:1} 

QCSE quenching phenomena similar to those just described have been
observed by Takeuchi {\it et al.} \cite{takeuchi} in InGaN wells,
with estimated fields in the 1 MV/cm range. The red shift and
optical inefficiency can in fact be removed, although only in a
transient fashion, by sufficiently high excitation powers.
The order of magnitude of the values reported in Ref.\onlinecite{takeuchi} is
$\sim$200 kW/cm$^2$ for 5 to 10 MQW periods, i.e. 20 to 40 kW/cm$^2$
per well, in qualitative agreement with our estimate above. One important
remark at this point is  that,  depending on the   excitation power,  
 the MQW will adsorb radiation at  many different
transition energies ranging from that of the built-in-field--biased
well (low power limit) to the quasi--flat-band well (high power limit)
-- that is, the MQW is a multistable switch. It is indeed fortunate
that the typical fields in these structures are such that one can 
physically access  the various possible regimes.

Another noticeable effect is that at a properly chosen value of the
sheet density (i.e. of the excitation power) one can obtain at the
same time a reasonable transition probability {\it and} a red-shifted
energy by just  increasing the well thickness. This is very useful since
the transition wavelength can be shifted to a different color without
changing alloy composition, but only the well thickness. For instance
(see Fig.\ref{fig.high.2}),
changing the well thickness from 20 to 30 {\AA} at a sheet density of
4$\times$10$^{12}$ cm$^{-2}$,  one obtains an energy shift of
0.1 eV at the cost of a
 loss of a factor 10 in recombination rate, which may still be acceptable
depending on the application. Red-shifting the transition energy in
this fashion may avoid the need to add e.g. some In in the QW
composition.

\subsection{Screening of fields in massive samples}
\label{sec.mass}

Free charge produced by  high excitation  screens polarization fields
fairly efficiently over the quite short distances typical 
 in nanostructures, since the spatial extension of the screening charge
is comparable to the size of the system. How about extended samples,
especially if not subjected to  illumination, i.e. having only intrinsic free
carriers ? 
It is indeed the case that no macroscopic
 fields exist in ``infinitely large'' samples even in the absence of high densities
 of (say) photogenerated carriers. The simple  reason is that the
 intrinsic carrier fluctuations in an undoped semiconductor rise
 exponentially as a function of deviations of the chemical potential
 from the  mid-gap value.\cite{mermin} In polarized  nitrides, such
 deviations occur due to the built-in fields. As the sample thickness increases,
 the potential drop grows linearly. When the drop is smaller than the
 gap, the field is uniform: $|\Evec|=4\pi\Pvec/\varepsilon_0$. 
 When the drop approaches  the gap value, i.e. for thicknesses approaching
 $d_c=E_{\rm gap}/|\Evec|$, the Fermi level nears the band edges:
 consequently, large amounts of  holes and electrons 
 are generated on the opposite sides of the sample. They
 screen partially the polarization charges, preventing the gap from closing.
 The total potential drop is thus  pinned at the gap value for all  thicknesses
$d>d_c$ -- that is, the effective gap decreases down to zero, but not 
below. For $d>d_c$,  the field will decrease  as
 $$|\Evec|=E_{\rm gap}/d.$$  For this picture to hold, the spatial
extension of the  screening charge at the sample surface  must be
comparable with that of  the polarization charge (which is a few {\AA}
at most \cite{noi.pol}) and much smaller that the sample size. This  
 will cause  the field  inside the sample to remain uniform, since the net
 effect of screening will be to change the effective polarization
 charge. In fact, this assumption turns out to be verified in
 practice on direct inspection, as we discuss below.

Clearly,   the above mechanism will strongly influence QW's of 
 thicknesses equal to, or larger than, the critical value $d_c$.
For the  QW we are considering here, with a polarization field
of --2.26 MV/cm, the critical value is $d_c \sim$ 165 {\AA}.
To confirm our picture, we  simulated  QW's with the same composition and
geometry considered in Sec. B, and thicknesses below and above 
 $d_c$, to mimic the crossover
from a ``microscopic'' to a ``macroscopic'' sample.  
In this case, we need to describe very extended bulk regions
on  the left and right of the QW, in order to account for the large
 screening length. Thus, we have made use of a classical Thomas-Fermi
 model where the charge densities are calculated with
 Fermi-Dirac statistics of a classical  system rather than  by solving
 the Schr\"odinger equation in the TB basis. 
This allows to consider devices with a spatial extension of several
hundreds of microns. Effective masses in this model where fitted with
the TB model in order to reproduce quite accurately the self-consistent 
TB results.
 
\begin{figure}
\epsfclipon
\epsfysize=7cm\centerline{\epsffile{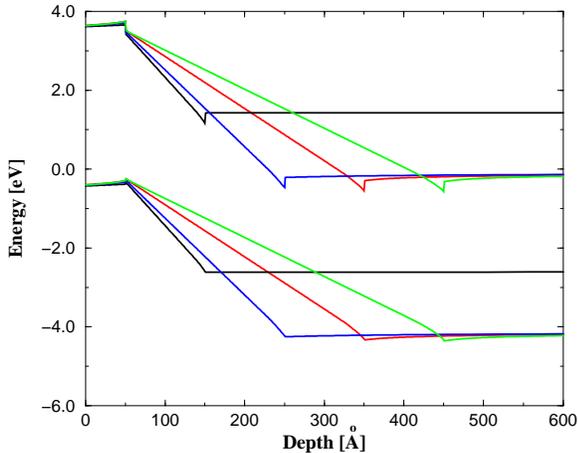}}
\caption{Reduction of polarization field in a thick layer near and
above the critical thickness.}
\label{fig.mass.1}
\end{figure}

The resulting 
self-consistent potential is shown in Fig.\ref{fig.mass.1} for 
well thicknesses of 100, 200, 300, and 400 {\AA}.
 lengths. A first point to note is that the field remains uniform for 
all well thicknesses. The field value
equals the polarization field 
for the smallest thickness (smaller than $d_c$); for the thicker
wells, the field (while  remaining uniform) indeed decreases
progressively as $\propto 1/d$. 

Photoluminescence experiments are not expected to be able to reveal
this effect (which should cause a saturation of the red shift 
as function of thickness) in very thick QW's. since the effective
recombination rate is vanishingly small at the relevant thicknesses.
Experiments aiming to reveal this effect should be
designed considering   our results that a very thick layer 
is effectively subjected to a uniform electrostatic field
$E_{\rm gap}$/$d$.In an unstrained GaN QW, for
$d>d_c$ (the latter being typically of order a few hundred {\AA} or
so depending on the polarization) the field is   $\sim$3.4 V/$d$, i.e.
 $\sim$70 kV/cm for $d=$0.5 $\mu$m. This is presumably sufficient to cause  
observable bulk-like effects such as  shifts in response functions or field
 effects on impurities.

A similar ``self-screening'' behavior has been  revealed
 indirectly   in devices comprising sufficiently thick layers. In
Ref. \onlinecite{yu} a 300 {\AA} thick  Al$_{0.15}$Ga$_{\rm 
0.85}$N layer was grown on a very thick GaN substrate, and topped with
a Schottky contact. The predicted field in the AlGaN layer is 1.4
MV/cm, which would cause 
across the layer a  potential drop 
of 4.2 eV.  The  maximum reasonable potential
drop dictated by 
Schottky barriers, conduction offset,  and Fermi level  is about  1
eV, so it must be the case that the polarization charge gets largely
screened by electrons from the GaN layer,  forming a high-density
two-dimensional electron gas (2DEG) at the heterointerface; this, by
the way, causes an enhanced mobility in the  conducting channel. CV
depth profiling indeed reveals a 2DEG at the interface.\cite{yu}
An equivalent, more formal description is: the  field, if assumed uniform,
 would force the metal-determined Fermi level to some 3 eV
above the conduction band of GaN, thus attracting towards the
interface  an enormous carrier density, which  screen (part of)
 the field out.   
Note in passing that in Ref. \onlinecite{yu} only piezoelectric
polarization  was considered, which leads to an underestimation of the
2DEG density, since the piezoelectric contribution is actually 
about one third of the total interface charge. 
Similar
considerations apply to other similar recent
experiments.\cite{bykhovski1}  A recent device simulation \cite{vogl}
has corrected this point, including in part the
spontaneous-polarization interface charges.  

\subsection{Suppressing QCSE by doping}

We have seen in the previous Sections that polarization fields can be
screened to a reasonable extent by generation of  free charge of
{\it both} kinds in the QW  upon e.g. optical excitation.
Qualitative problems with this screening mechanism are
that {\it (a)}  it is transient, since it disappears when photoexcitation
or current injection are removed, and that
{\it (b)} in purely electronic (i.e. non-optoelectronic) devices,
it is unlikely that the high densities needed can be reached in normal
operating conditions. Besides, the injected current is not constant
in time, so that the well shape changes in time.

\begin{figure}
\epsfclipon
\epsfysize=7cm\epsfxsize=9cm\centerline{\epsffile{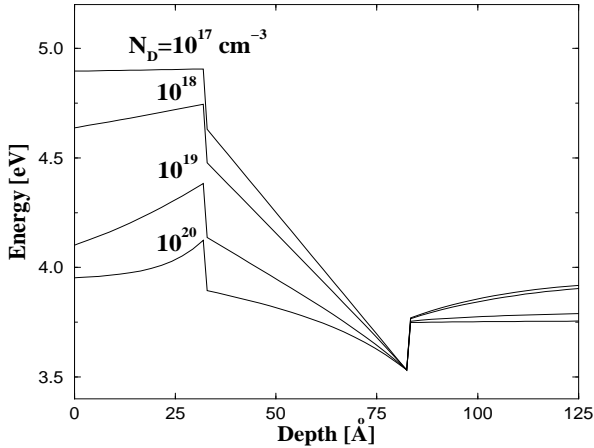}}
\caption{Selfconsistent conduction band edge of a remotely doped  well.}
\label{fig.dope}
\end{figure}

It is natural to presume that the same effects can be achieved in a
permanent fashion using extrinsic carriers from dopants. 
The idea is to provide the well
with carriers which would screen the polarization charge, excepts that
now  the electrons are released into the QW from the doped barriers,
and not injected or photogenerated. 
Of course, this effect is not
transient as the others discussed previously.  
The problem
is, how high must the doping density be to achieve the same level of
screening as in a high optical excitation regime.
We simulated 
a 50 {\AA} thick Al$_{0.2}$Ga$_{0.8}$N/GaN single QW,
where the barriers have been doped $n$- type in the range from 
10$^{17}$ to 10$^{20}$ cm$^{-3}$ and the donor ionization energy have 
been set to 10 meV;\cite{aln} 
we used in this 
simulation a selfconsistent TB approach. 
The resulting conduction band profile 
is displayed in Figure \ref{fig.dope} for the various
doping densities. 
%
The polarization field raises the conduction band on the left side 
over the Fermi energy  and 
in order for the barrier conduction band to reach the Fermi level on
the far left,
the electrons are transferred from the left-side barrier into the QW,
leaving behind a large depletion layer.
As a consequence of the electron flow into the QW, the polarization field
start to get significantly screened at doping densities  
above $\sim$ 10$^{19}$cm$^{-3}$. 
The existence of a depletion layer causes 
a large band bending in the barrier, while the bending is absent 
in the left half of the well. This is  quite different to the case 
of the photoexcited well, where the bending on the left side of the 
well was due to hole accumulation 
nearby the interface (see Fig. 6).
This explains why the field remains nearly uniform 
in the left half of the well for all of the simulations performed.
On the right side of the well only a small bending due to the electron 
accumulation is present.
Indeed electron localization is quite weak in these systems since 
the confinement potential is 
small as compared to the field-induced drop, and electrons
tend to spill over to the right-side barrier. This is
likely to be case in all nitride systems in this composition range.

From these results, we conclude that doping can indeed
be used to screen polarization fields. While  
it is not obvious that the needed doping level can always be reached in
practice, it is likely  
that a combination of doping and 
current injection or photoexcitation
will generally succed in quenching the polarization field 
in the range of MV/cm 
recovering a quasi flat band condition. 
Fields in  InGaN/GaN systems will be
generally smaller than  
those in AlGaN/GaN systems for typical compositions in use today,
and will therefore be more easily amenable to treatment  by 
the above technique.
This  procedure has in fact been adopted in experiment by
Nakamura's group,\cite{nak} which  reported that a doping level of 10$^{19}$
cm$^{-3}$  is sufficient to quench the QCSE to a large extent.
Indeed in their In$_{0.15}$Ga$_{0.85}$N/GaN MQW's 
the unscreened field is  $\sim$1.2 MV/cm, i.e. approximately a half of
the one we considered here thus easily screened by remote 
doping, this conclusion is in qualitative 
agreement with our findings. 


\section{Summary and acknowledgements}
In conclusion, we have discussed how macroscopic (and in particular,
spontaneous) polarization 
plays an important  role in nitride-based MQW's by producing large
built-in electric fields. Contrary to zincblende  semiconductors, in
III-V nitrides--based devices the spontaneous polarization is an
unavoidable source of large electric fields even in lattice-matched
(unstrained) systems.   The existence of these fields may also be used
as additional degree of freedom in device design: for instance, for
an appropriate choice of alloy composition, spontaneous and
piezoelectric fields may be caused to cancel out, thus freeing the
structures from built-in fields.
We have also discussed the different regimes of  free carrier
screening, effected by doping or optical excitation, showing that 
fields can be screened  only in the presence of high free carrier
densities,
which leads to unusually high lasing thresholds for undoped QW's.
Of course, our results about the effects on the electronic structure
apply qualitatively to any kind of polarization field, thus in
particular also to piezo-generated  ones.

VF and FB acknowledge special support from the PAISS program of
 INFM. VF's stay at the Walter Schottky Institut  was supported by the
 Alexander von Humboldt-Stiftung. FDS, ADC and PL acknowledge support from 
Network Ultrafast and 40\% MURST.


\end{multicols}

\begin{thebibliography}{99}
\bibitem{nota3}
>From dielectric displacement conservation (see Ref.\onlinecite{noi.eps})
 one gets $4\pi (P_{\rm C}-P_{\rm A}) =
\varepsilon_{\rm A} E_{\rm A} -\varepsilon_{\rm C} E_{\rm
C}$. Periodicity implies  $\ell_{\rm A} E_{\rm A} + \ell_{\rm C}
E_{\rm C} =0$. Solving for  one of the fields, 
Eqs.\ref{eq.piezo} are immediately obtained.
\bibitem{nota1}
This assumption is valid in the limit in which the electric field  
across the whole structure is negligible 
with respect to the internal fields induced by polarization.  This is 
verified e.g. when the MQW system is embedded on both sides by 
doped materials of the same composition.
\bibitem{nota}
Note that for normal devices layered along the nominal 
growth direction, the electric field is always parallel to the growth 
direction,  even if the polarization is not.
\bibitem{nye2} 
One possible way to see it is as the
zero-strain limit of the piezoelectric 
polarization.
\end{thebibliography}
\end{document}